\documentclass[%
reprint,
superscriptaddress,
nobibnotes,
amsmath,amssymb,
aps,longbibliography,
prl,
floatfix,
]{revtex4-2}

\usepackage[T1]{fontenc}
\usepackage[utf8]{inputenc}
\usepackage[english]{babel}
\makeatletter
\@ifundefined{selectlanguage}{
}{
  \expandafter\newlanguage\csname l@en\endcsname %
}
\makeatother
\usepackage{graphicx}%
\usepackage{verbatim}
\usepackage[version=4]{mhchem}
\usepackage{xcolor}
\usepackage{physics}
\usepackage{comment}
\usepackage[normalem]{ulem}
\usepackage{pdfpages}
\usepackage{pgffor}
\usepackage{booktabs}
\usepackage{makecell}
\usepackage{float}                      %

\newcommand{\rev}[1]{#1}

\usepackage{etoolbox}                   %
\usepackage{caption}                    %

\AtBeginEnvironment{listing}{%
  \raggedright
  \captionsetup{justification=raggedright,singlelinecheck=false}%
}

\usepackage[normalem]{ulem}
\definecolor{tangerine}{rgb}{0.944,0.522,0}
\definecolor{verde}{rgb}{0.,0.6,0}
\definecolor{rosso}{rgb}{0.9,0.0,0.2}
\definecolor{orange}{rgb}{1.0,0.5,0.0}

\newif\ifhighlight

\newcommand{\highlight}{\highlighttrue}
\highlight
\newcommand{\editor}[2]{%
  \expandafter\newcommand\csname #1note\endcsname[1]{%
    \textcolor{#2}{(\textbf{#1note:} \textsc{##1})}}%
  \expandafter\newcommand\csname #1\endcsname[1]{%
    \ifhighlight\textcolor{#2}{##1} \else ##1\fi}%
  \expandafter\newcommand\csname #1cancel\endcsname[1]{%
    \ifhighlight\textcolor{#2}{\sout{##1}}\fi}%
  \expandafter\newcommand\csname #1change\endcsname[2]{%
    \ifhighlight\textcolor{#2}{\sout{##1} ##2}\else ##2\fi}%
  \newenvironment{#1text}{\ifhighlight\color{#2}\fi}{\color{black}}
}

\editor{MC}{red}
\editor{MK}{orange}
\newcommand{\grayout}[1]{}

\newcommand{\shiftmltwo}{ShiftML2}
\newcommand{\shiftmlthree}{ShiftML3}
\newcommand{\shiftmlthreeshort}{sML3}
\newcommand{\shiftmltwoshort}{sML2}

\makeatletter
\AtBeginDocument{\let\LS@rot\@undefined}
\makeatother

\begin{document}

\title{A deep learning model for chemical shieldings\\in molecular organic solids including anisotropy }

\author{Matthias Kellner}
\affiliation{Laboratory of Computational Science and Modeling, Institut des Mat\'eriaux, \'Ecole Polytechnique F\'ed\'erale de Lausanne, 1015 Lausanne, Switzerland}
\author{Jacob B. Holmes}
\affiliation{Laboratory of Magnetic Resonance, Institut des Sciences et Ing\'enierie Chimiques, \'Ecole Polytechnique F\'ed\'erale de Lausanne, 1015 Lausanne, Switzerland}
\author{Ruben Rodriguez-Madrid}
\affiliation{Laboratory of Magnetic Resonance, Institut des Sciences et Ing\'enierie Chimiques, \'Ecole Polytechnique F\'ed\'erale de Lausanne, 1015 Lausanne, Switzerland}
\author{Florian Viscosi}
\affiliation{Laboratory of Magnetic Resonance, Institut des Sciences et Ing\'enierie Chimiques, \'Ecole Polytechnique F\'ed\'erale de Lausanne, 1015 Lausanne, Switzerland}
\author{Yuxuan Zhang}
\affiliation{Laboratory of Computational Science and Modeling, Institut des Mat\'eriaux, \'Ecole Polytechnique F\'ed\'erale de Lausanne, 1015 Lausanne, Switzerland}
\author{Lyndon Emsley}
\affiliation{Laboratory of Magnetic Resonance, Institut des Sciences et Ing\'enierie Chimiques, \'Ecole Polytechnique F\'ed\'erale de Lausanne, 1015 Lausanne, Switzerland}
\email{lyndon.emsley@epfl.ch}
\author{Michele Ceriotti}
\email{michele.ceriotti@epfl.ch}
\affiliation{Laboratory of Computational Science and Modeling, Institut des Mat\'eriaux, \'Ecole Polytechnique F\'ed\'erale de Lausanne, 1015 Lausanne, Switzerland}

\newcommand{\CSM}[1]{{\color{red}#1}}

\date{\today}%

\begin{abstract}
\textbf{Abstract}: Nuclear Magnetic Resonance (NMR) chemical shifts are powerful probes of local atomic and electronic structure that can be used to resolve the structures of powdered or amorphous molecular solids. 
Chemical shift driven structure elucidation depends critically on accurate and fast predictions of chemical shieldings, and machine learning (ML) models for shielding predictions are increasingly used as scalable and efficient surrogates for demanding ab initio calculations. However, the prediction accuracies of current ML models still lag behind those of the DFT reference methods they approximate.
Here, we introduce \shiftmlthree{}, a deep-learning model that improves the  accuracy of predictions of isotropic chemical shieldings  in molecular solids, and does so while also predicting the full shielding tensor. 
On experimental benchmark sets, we find root-mean-squared errors with respect to experiment for \shiftmlthree{} that approach those of DFT reference calculations, with RMSEs of 0.53 ppm for $^{1}$H, 2.4 ppm for $^{13}$C, and 7.2 ppm for $^{15}$N, compared to DFT values of 0.49 ppm, 2.3 ppm, and 5.8 ppm, respectively.

\end{abstract}

\maketitle

Chemical function is directly related to structure. As a result, the determination of atomic-level three-dimensional structures is at the heart of molecular and materials science. In this context, solid-state NMR (often in conjunction with diffraction methods and other spectroscopic methods) has emerged as the method of choice for atomic-level characterization of complex materials in powder form. 
This has been driven by the development of  methods to probe the local atomic environment based on chemical shifts,~\cite{lawsSolidStateNMRSpectroscopic2002a, ashbrookCombiningSolidstateNMR2016, hodgkinsonNMRCrystallographyMolecular2020,reifSolidstateNMRSpectroscopy2021a, emsleySpiersMemorialLecture2025, bryce_modern_2025} often referred to as NMR crystallography. 
In turn, this has been made possible by the development of accurate density functional theory (DFT) based methods to calculate chemical shifts.~\cite{pickardAllelectronMagneticResponse2001a, hartmanBenchmarkFragmentbased12016} Indeed, it has been recently shown that NMR can elucidate structures of molecular solids from powders with resolution better than 0.2 Å.~\cite{hofstetter_positional_2017, holmesImagingActiveSite2022}

In a nutshell, NMR crystallography consists of experimentally measuring and assigning chemical shifts for each of the atomic sites in a structure and then comparing the measured shifts to those predicted by computation for a pool of candidate structures, that are themselves enumerated by a computational search. 
Chemical shifts  are traditionally calculated using DFT methods.~\cite{pickardAllelectronMagneticResponse2001a, yatesCalculationNMRChemical2007, bonhommeFirstPrinciplesCalculationNMR2012, beranModelingPolymorphicMolecular2016, ramosInterplayDensityFunctional2024}
This workflow has been successful for materials ranging from hybrid-perovskites~\cite{hope_nanoscale_2021, kubickiNMRSpectroscopyProbes2021} and inorganic materials~\cite{senkerMicroscopicDescriptionPolyamorphic2005, leeStructureDisorderAmorphous2010, grasCrystallineAmorphousCalcium2016, walkleySolidstateNuclearMagnetic2019, kunhi_mohamed_atomic-level_2020, morales-melgares_atomic-level_2022} to enzyme active sites,~\cite{singh_fast_2019, bertarelloPicometerResolutionStructure2020, holmesImagingActiveSite2022}  organic molecular solids,~\cite{middletonMolecularConformationsPolymorphic2000, yatesCombinedFirstPrinciples2005, harrisAssigningCarbon13NMR2006, harrisNMRcrystallographyOxybuprocaineHydrochloride2007, salagerPowderCrystallographyCombined2010, harrisComputationNMRCrystallography2010, baiasPowderCrystallographyPharmaceutical2013, baiasNovoDeterminationCrystal2013, pinonPolymorphsTheophyllineCharacterized2015, pindelskaSolidStateNMREffective2015, rezendeCombiningNuclearMagnetic2016, tianApplicationSolidStateNMR2018, szeleszczukApplicationCombinedSolidstate2019, daiSolidState1H13And172020, southernAnnualReportsNMR2021} and notably for amorphous drug formulations.~\cite{cordovaStructureDeterminationAmorphous2021a, holmesAtomiclevelStructureAmorphous2024, guestEssentialSynergyMD2025, torodiiThreeDimensionalAtomicLevelStructure2025} One of the key bottlenecks to address in this workflow is both the computational cost and the accuracy of chemical shift predictions. 

The introduction of accurate machine learning models to predict chemical shifts in solids has essentially removed the computational cost barrier, allowing for tens of millions of shifts to be predicted on the order of days.~\cite{cunyInitioQualityNMR2016, paruzzo_chemical_2018, liuMultiresolution3DDenseNetChemical2019a, chakerNMRShiftsAluminosilicate2019, unzueta_predicting_2021, cordova_machine_2022, liHighlyAccuratePrediction2024,  venetosMachineLearningFull2023, mahmoudGraphneuralnetworkPredictionsSolidstate2024, liHighthroughputCalculationsMachine2025} 
For organic molecular solids, Paruzzo \textit{et al.}~\cite{paruzzo_chemical_2018} developed a model (dubbed ShiftML) that could predict isotropic chemical shifts for any molecular solid containing C, H, N, O atoms, that was trained on GIPAW DFT data for 2,000 structures from the Cambridge Structural Database (CSD).~\cite{groomCambridgeStructuralDatabase2016} Engel and coworkers later broadened its applicability to sulfur-containing compounds.~\cite{engelBayesianApproachNMR2019}
The model was subsequently improved and extended to \shiftmltwo{}~\cite{cordova_machine_2022}  which was trained on an extended set of over 14,000 structures containing H, C, N, O, S, F, P, Cl, Na, Ca, Mg and K, and composed of both relaxed and thermally perturbed structures. Both ShiftML models used kernel ridge regression (KRR),~\cite{murphyMachineLearningProbabilistic2014} with uncertainties in the predictions estimated using a committee of 32 models (for \shiftmltwo{}).~\cite{musi+19jctc} The local atomic environments were described using smooth overlap of atomic positions (SOAP) descriptors.~\cite{bartokRepresentingChemicalEnvironments2013}

The predictions of \shiftmltwo{} yield an accuracy for isotropic \textsuperscript{1}H shifts of 0.5~ppm with respect to the DFT computed shifts, and an accuracy of 0.47~ppm with respect to a benchmark set of experimental shifts.~\cite{cordova_machine_2022} Increasing the accuracy of shift prediction for ML models would directly translate to better structure determination. Indeed, Unzueta \textit{et al.}~\cite{unzueta_predicting_2021} have shown that using a $\Delta$-ML approach to improve the prediction accuracy of DFT computed shifts themselves, by correcting the effects of small basis set sizes in DFT shielding calculations, can lead to errors as small as 0.11~ppm for \textsuperscript{1}H shifts with respect to DFT. Although this is not directly relevant to pure ML predictions, since they use DFT computations as a starting point, it does nicely illustrate the potential to develop more accurate models.

Even though DFT calculations naturally yield the full chemical shielding tensor, the prediction models developed so far for molecular solids were only trained on isotropic values of the chemical shift (where the isotropic shift is the average of the three principal components). Recently, models for silicon oxides~\cite{venetosMachineLearningFull2023, mahmoudGraphneuralnetworkPredictionsSolidstate2024} and non-magnetic oxides~\cite{liHighthroughputCalculationsMachine2025}  have been introduced to predict the full chemical shift tensor with orientation, extending the scope of ML-predicted chemical shieldings to applications that rely on the chemical shift anisotropy (CSA).
Tensorial models have also been used to predict other quantities relevant to nuclear spectroscopy, such as electric field gradients \cite{f.harperTrackingLiAtoms2025a} and zero-field splitting tensors \cite{zaverkinThermallyAveragedMagnetic2022c}.

Here, we present a new machine learning model (dubbed \shiftmlthree{}) that predicts the full chemical shielding tensors, trained using shieldings calculated at the GIPAW/PBE level of theory. 
The model is based on an ensemble of \textit{Point Edge Transformer} (PET) models.~\cite{NEURIPS2023_fb4a7e35} The model not only predicts anisotropic shieldings, but also yields significantly improved accuracy with respect to DFT, with up to a two-fold improvement over \shiftmltwo{} for the isotropic shieldings of \rev{nuclei with Z~\textgreater~1}.

Isotropic shifts are often sufficient to identify the experimentally-observed crystal structure out of a pool of potential candidates. For this reason, previous ShiftML models have been constructed by directly learning the isotropic chemical shieldings $\sigma_{\text{iso}}$ from DFT calculations, using symmetry-adapted models designed to make predictions that are exactly invariant to rigid rotations of a structure. However, the chemical shielding is a tensorial property, whose anisotropy (typically referred to as the chemical shift anisotropy or CSA) provides additional information that can be useful to explore crystal structures.~\cite{harperMeasuringModelingAnisotropy2024}
Given that GIPAW-DFT calculations yield the full chemical shielding tensor, in \shiftmlthree{} we include tensorial predictions, focusing on the symmetric component which is of experimental relevance. For molecular solids, isotropic chemical shifts are readily measured using fast magic-angle spinning (MAS) methods,~\cite{reifSolidstateNMRSpectroscopy2021a} and CSA can be measured using slow MAS~\cite{herzfeldSidebandIntensitiesNMR1980} or using multi-dimensional isotropic-anisotropic correlation experiments to avoid spectral crowding.~\cite{dixonSpinningsidebandfreeSpinningsidebandonlyNMR1982,  ganHighresolutionChemicalShift1992, gannDynamicangleSpinningSidebands1993, antzutkinTwoDimensionalSidebandSeparation1995} 
While these parameters are readily measurable, there is currently no sufficiently large database of experimental shifts for molecular solids with corresponding high-resolution atomic-level structures. This is in contrast to small molecules and proteins in solution, which have large and well-curated experimental databases.~\cite{meilerPROSHIFTProteinChemical,  hanSHIFTX2SignificantlyImproved2011} 
As a result, ML models in solids have been developed by using curated sets of structures together with shifts computed using DFT. 

Here we use the training set from the most recent previous ShiftML model, and for which the computed shifts (computed at the PBE level of theory) contain the full tensor information.~\cite{cordova_machine_2022} 
This dataset is composed of molecular solids where the structures include (up to) 12 elements [H, C, N, O, F, Na, Mg, P, S, Cl, K, Ca] commonly found in organic compounds. It contains a total of   14,146  structures, including 7,548 sampled from the Cambridge Structural Database (CSD) via farthest point sampling (FPS)~\cite{imba+18jcp} to cover a broad range of structural motifs, and 6,598 additional thermally distorted structures. 
The \shiftmltwo{} test set was divided into a new validation (505 structures, with 41,599 atomic environments) and a new test set (1,254 structures, with 106,230 atomic environments). 
In this split we chose a stratified selection strategy that ensures that both the distorted and the relaxed structure of any one CSD identifier are found either in the validation or the test set, avoiding data leakage between the sets. Furthermore, we ensure that both the new validation and new test sets contain at least a few environments of all 12 chemical elements.  
To compare the accuracies between \shiftmltwo{} and \shiftmlthree{} in a consistent fashion,  the accuracies are reported on this new test set (referred to as the \emph{CSD-test} set). 
More details of the training, validation and test sets for each element in \textit{CSD-test} are given in section~\ref{sec:detailed_preds_sML5} of the supporting information.
Due to computational constraints, \shiftmltwo{} was trained on a representative sub-selection of atomic environments from the training set. The favorable computational scaling of neural network training via stochastic gradient descent allowed us to train \shiftmlthree{} using all 1.4 million atomic environments contained in the 14,146 structures in the training set.
The updated dataset containing the chemical shielding tensor data is available via a public data record associated with this publication.~\cite{kellner2025deep}
We also use established datasets of experimental \rev{isotropic} chemical shielding values,  with 13 structures for $^{1}$H, 21 for $^{13}$C, and 15 for $^{15}$N,~\cite{cordova_machine_2022, ramosInterplayDensityFunctional2024} and a benchmark set of experimental $^{13}$C CSA principal values for 19 compounds.~\cite{hartmanFragmentbased13CNuclear2015}

For the construction of the \shiftmlthree{} model, we learn the correlation between chemical shielding tensor and structure using an ensemble of \textit{Point Edge Transformer} (PET) models.~\cite{NEURIPS2023_fb4a7e35} 
PET is an unconstrained graph neural network (GNN) architecture that delivers accurate and fast predictions for several material properties benchmark sets, importantly also on rotationally equivariant targets such as molecular dipole moments.~\cite{NEURIPS2023_fb4a7e35}
For \shiftmlthree{}, we use a modified version of the PET architecture referred to as nanoPET (available via the \texttt{metatrain} library).~\cite{MetatensorMetatrain2025}
Even though nanoPET does not enforce rotational invariance or equivariance of the target predictions, it learns the rotational behaviour of the targets via rotational augmentation during the training process -- a mechanism that has been shown to yield models that are symmetry-compliant to a high accuracy.~\cite{langerProbingEffectsBroken2024} 

In practice, nanoPET predicts the chemical shielding tensors in their irreducible spherical tensor (IST) representation in a manner similar to symmetry-adapted, rotationally-constrained models.~\cite{gris+18prl} 
The chemical shielding tensors are decomposed into their irreducible real spherical representations: A scalar, $\lambda=0$ part, a $\lambda=1$ pseudovector, and a $\lambda=2$ proper 5-vector having the same rotational behaviour as their corresponding $Y^{\lambda}$ spherical harmonics.~\cite{haeberlenHighResolutionNMR1976}
We refer the reader to the following references~\cite{anetShieldingTensorPart1991, facelliChemicalShiftTensors2011, youngTensorViewSoftwareTool2019} for a more in-depth discussion of the properties of the chemical shielding tensor and a discussion of its rotational properties. 
nanoPET uses a single message-passing sequence of unconstrained graph transformer modules, with one final linear readout layer per IST component, $T^{\lambda}$.  
The final tensor predictions are then obtained by a change of representation back from the predicted IST components to the tensor in the Cartesian basis.

During the training procedure, random rotations in their irreducible Wigner-D matrix representations are automatically applied to the target IST components at each training step, as well as random inversions. 
We train an ensemble consisting of 8 nanoPET models. The models are trained for 1,500 epochs, with a constant learning rate of $3\cdot 10^{-4}$. We train the model using mini-batch gradient descent with the Adam optimizer~\cite{kingmaAdamMethodStochastic2017a} and a batch size of 4. 
The construction of an ensemble model has several benefits; in addition to increased prediction accuracies, an ensemble also provides uncertainty estimates and further reduces the rotational symmetry breaking of the nanoPET predictions.~\cite{misofEquivariantNeuralTangent2025, gerkenEmergentEquivarianceDeep2024}
In Section~\ref{sec:SI_rotationa_fluc} of the supporting information we make a detailed assessment of the residual symmetry errors of the \shiftmlthree{} model.

When performing hyperparameter optimization we observed that the prediction accuracies of individual nanoPET models do not vary strongly between reasonable guesses of potential hyperparameter choices (cutoffs, number of message passing layers and token size). 
We therefore employ a set of hyperparameters for the nanoPET implementation that provides a good compromise between accuracy and evaluation speed. An in-depth evaluation of the computational speed of the \shiftmlthree{} model can be found in the section~\ref{sec:SI_computational_timing} of the supporting information.
We choose a radial cutoff of 5 Å to encode local information (displacement vectors, atom types and messages from neighbouring atoms) into tokens. \shiftmlthree{} employs 2 message passing layers and 2 attention layers per message passing step, and a token dimensionality of 192. For better usability, we make \shiftmlthree{} available through a Python package (more information in section~\ref{SI:implementation} of the supporting information).
The predictions are affected by a structure-dependent uncertainty, which should be quantified to assess how reliably the model can be used for a structure determination task -- especially to identify the degradation of accuracy that usually occurs in the extrapolative regime. 
In many cases the uncertainty can be attributed to lack of knowledge of the model due to improper sampling of the training data (epistemic uncertainties).~\cite{hullermeierAleatoricEpistemicUncertainty2021c} %
 Here, we refer to the prediction accuracy of the model with respect to the DFT calculations, whereas the underlying error of the electronic-structure reference would need to be assessed by comparison with experiments or higher-level calculations. 
PET and most other atomistic neural networks make only point estimates of the target quantity. Constructing an ensemble is a universal strategy to equip ML models with uncertainty estimates,~\cite{lakshminarayananSimpleScalablePredictive2017a, rafteryUsingBayesianModel2005a} even if the model does not allow for a probabilistic interpretation of its predictions. 
An ensemble model is constructed by training multiple models. The final prediction for an atomic local environment, ${A}_i$, of an ensemble of N$_{ens}$ models, is then taken to be the mean $\bar{y}(A_i)$ of the individual predictions $y^{k}(A_i)$ of the ``committee'' members,  and their spread is taken to be an indicator of the reliability of the model predictions:
\begin{equation}
    \label{eq:UQ}
    \sigma^{2}_{\text{pred}}(A_i) = \frac{1}{N_{\text{ens}}-1} \sum_{k=1}^{N_{\text{ens}}} \big[y^{k}(A_i) - \overline{y}(A_i) \big]^2
\end{equation}
 We compute uncertainty estimates from the committee of PET model predictions.
We note in passing that it is often possible to reduce the overhead of ensemble models by building \emph{shallow ensembles} that share the time-consuming part of the evaluation, and differ only by their last-layer features.~\cite{kellnerUncertaintyQuantificationDirect2024a}
Here we opt for a full ensembling to maximise prediction accuracies in line with previous works on prediction chemical shieldings with deep learning networks.~\cite{liuMultiresolution3DDenseNetChemical2019a, unzueta2022low} 
Following standard practice,~\cite{guoCalibrationModernNeural2017a, kuleshovAccurateUncertaintiesDeep2018c} we calibrate the uncertainty estimates of \shiftmlthree{} on the validation set, scaling the predicted ensemble uncertainties with a global, input-independent correction.~\cite{musi+19jctc} More details of the model calibration and evaluation of the uncertainty estimates can be found in section~\ref{sec:SI_UQ} of the supporting information.

\grayout{Within the Bayesian NMR crystallography framework\cite{engelBayesianApproachNMR2019, muellerUniformChisquaredModel2025} a total site-specific uncertainty could then, for example simply be obtained via Gaussian error calculus
(neglecting correlations between $\sigma^2_{\text{DFT,exp}}$ and
$\sigma^2_{\text{pred,ML}}$). 
\begin{equation}
    \sigma_{\text{total}} = \sqrt{\sigma^2_{\text{pred,ML}} + \sigma^2_{\text{DFT,exp.}}} 
\end{equation}
}

\begin{figure}[tbp]
    \centering
    \includegraphics[width=\columnwidth]{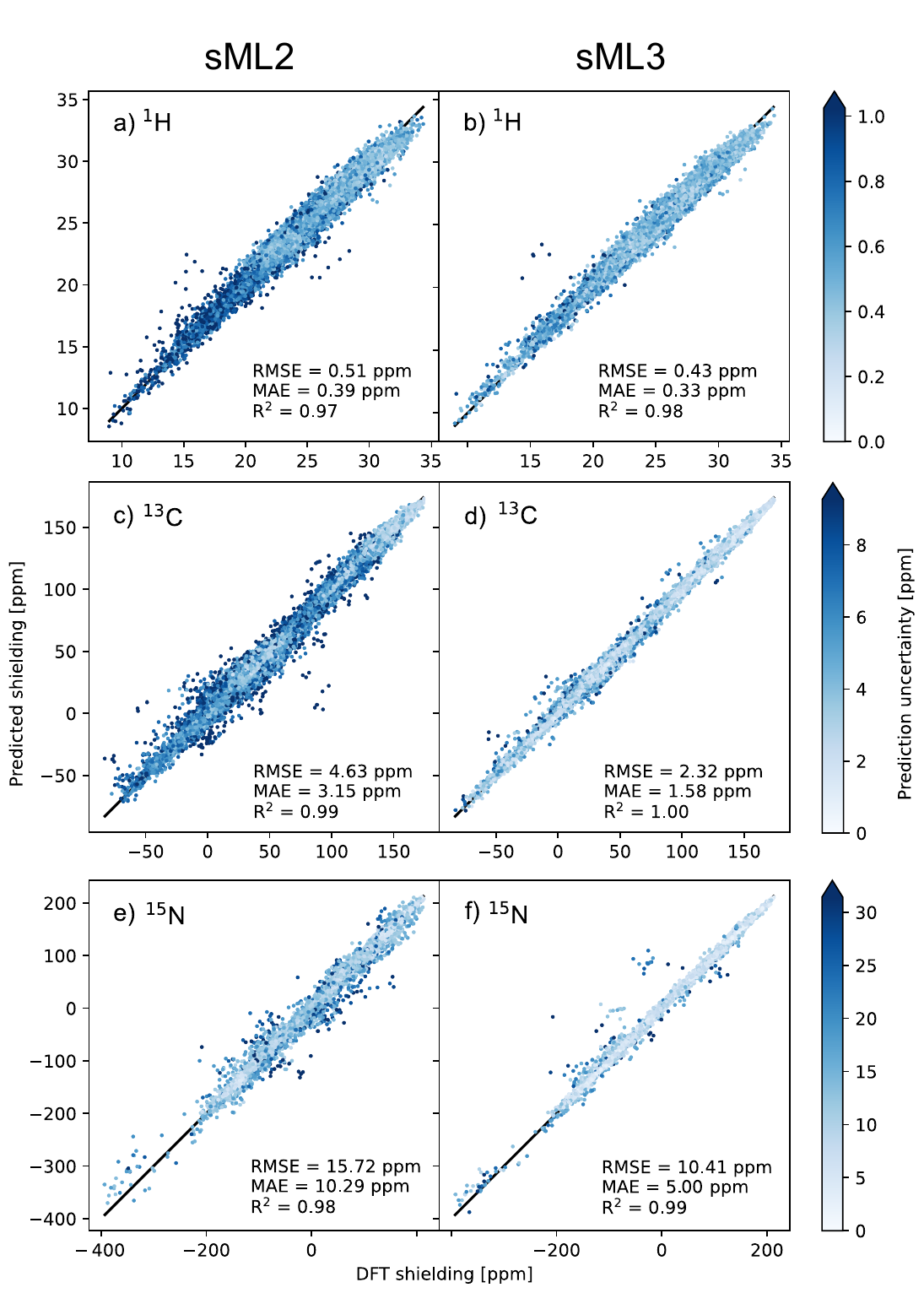}
    \caption{Parity plots between DFT-computed shieldings and predictions using \shiftmltwo{} or \shiftmlthree{} for (a,b) $^{1}$H, (c,d) $^{13}$C and (e,f) $^{15}$N evaluated on the new \textit{CSD-test} set. The black line shows perfect correlation. Each point is colored according to the rescaled prediction uncertainty associated with each atomic site.}
    \label{fig:parity_plot}
\end{figure}

To evaluate the performance of \shiftmlthree{}, we first compare predicted isotropic chemical shieldings against DFT calculated isotropic shieldings on the \textit{CSD-test} set. %
Figure~\ref{fig:parity_plot} shows parity plots illustrating the overall agreement between \shiftmlthree{} and DFT \rev{for the nuclei that are most often used in NMR crystallography. 
We observe an RMSE of 0.43~ppm, 2.3~ppm and 10.4~ppm for \textsuperscript{1}H, \textsuperscript{13}C and \textsuperscript{15}N, respectively.
These correspond to an error reduction relative to \shiftmltwo{} between 20\%  for \textsuperscript{1}H, to a factor of 2 for \textsuperscript{13}C.} 
Similar levels of improvement are seen also for almost all nuclei represented in the dataset, as shown in Table~\ref{tab:iso_accuracy} (with the exception of \textsuperscript{39}K, for which there are only 6 atomic environments in the test set, making the assessment inconclusive).

\begin{table}[tbp]
\caption{\label{tab:iso_accuracy} Prediction accuracies of \shiftmltwo{} (\shiftmltwoshort) and \shiftmlthree{} (\shiftmlthreeshort) models on the \textit{CSD-test} dataset. MAE and RMSE (in ppm) are evaluated on a hold-out test set.} %
\begin{tabular}{@{}>{\hspace{3 mm}}l @{\hskip 2mm}cc@{\hskip 4mm}cc@{\hskip 4mm}cc@{\hskip 2mm}}
\toprule
\multicolumn{1}{c@{\hskip 3mm}} {nucleus} & \multicolumn{2}{c@{\hskip 4mm}}{MAE } & \multicolumn{2}{c@{\hskip 4mm}}{RMSE} & \multicolumn{2}{c@{\hskip 2mm}}{$R^2$} \\
\midrule
 & \shiftmltwoshort & \shiftmlthreeshort & \shiftmltwoshort & \shiftmlthreeshort & \shiftmltwoshort & \shiftmlthreeshort \\
\midrule
$^{1}\text{H}$ & 0.39 & 0.33 & 0.51 & 0.43 & 0.97 & 0.98 \\
$^{13}\text{C}$ & 3.15 & 1.58 & 4.63 & 2.32 & 0.99 & 1.00 \\
$^{15}\text{N}$ & 10.29 & 5.00 & 15.72 & 10.41 & 0.98 & 0.99 \\
$^{17}\text{O}$ & 16.11 & 7.51 & 22.96 & 11.45 & 0.98 & 1.00 \\
$^{19}\text{F}$ & 7.77 & 4.42 & 10.83 & 6.66 & 0.97 & 0.99 \\
$^{33}\text{S}$ & 34.35 & 15.05 & 53.34 & 26.34 & 0.88 & 0.97 \\
$^{31}\text{P}$ & 21.83 & 9.49 & 39.20 & 18.98 & 0.62 & 0.91 \\
$^{35}\text{Cl}$ & 17.40 & 10.44 & 23.85 & 15.11 & 0.96 & 0.98 \\
$^{23}\text{Na}$ & 7.64 & 4.09 & 8.10 & 4.61 & 0.00 & 0.68 \\
$^{43}\text{Ca}$ & 5.81 & 2.96 & 6.98 & 3.34 & 0.81 & 0.96 \\
$^{25}\text{Mg}$ & 10.69 & 3.68 & 14.55 & 5.49 & 0.66 & 0.95 \\
$^{39}\text{K}$ & 3.42 & 4.29 & 4.28 & 4.65 & 0.77 & 0.73 \\
\bottomrule
\end{tabular}
\end{table}

\begin{table}[tbp]
\caption{\label{tab:cs_tensor_derived_values} Prediction accuracies of \shiftmlthree{} on symmetric tensor components $\sigma_{ij}$ and principal tensor components $\sigma_{\text{PAS}}$. 
MAE and RMSE (in ppm) are evaluated on a hold-out test set.}
\begin{tabular}{@{}>{\hspace{3 mm}}l @{\hskip 2mm}ccc@{\hskip 6mm}ccc}
\toprule
    \multicolumn{1}{c@{\hskip 3mm}} {nucleus} & \multicolumn{3}{c}{$\sigma_{ij}$} & \multicolumn{3}{c}{$\sigma_{\text{PAS}}$} \\
\midrule
 & MAE & RMSE & $R^2$ & MAE & RMSE & $R^2$ \\
\midrule
$^{1}\text{H}$ & 0.54 & 0.72 & 1.00 & 0.65 & 0.85 & 0.98 \\
$^{13}\text{C}$ & 2.10 & 3.08 & 1.00 & 2.64 & 3.88 & 1.00 \\
$^{15}\text{N}$ & 5.32 & 10.75 & 0.99 & 7.21 & 15.62 & 0.99 \\
$^{17}\text{O}$ & 7.58 & 12.09 & 0.99 & 10.28 & 16.79 & 1.00 \\
$^{19}\text{F}$ & 4.05 & 6.28 & 1.00 & 5.73 & 8.65 & 0.99 \\
$^{33}\text{S}$ & 17.15 & 30.37 & 0.97 & 22.53 & 42.73 & 0.97 \\
$^{31}\text{P}$ & 10.30 & 18.73 & 0.98 & 13.35 & 23.59 & 0.95 \\
$^{35}\text{Cl}$ & 10.14 & 15.32 & 1.00 & 14.02 & 20.50 & 0.99 \\
$^{23}\text{Na}$ & 2.83 & 4.01 & 1.00 & 4.28 & 5.24 & 0.91 \\
$^{43}\text{Ca}$ & 3.14 & 3.85 & 1.00 & 4.53 & 4.93 & 0.98 \\
$^{25}\text{Mg}$ & 7.48 & 9.79 & 1.00 & 6.27 & 8.34 & 0.93 \\
$^{39}\text{K}$ & 2.93 & 3.95 & 1.00 & 4.42 & 5.21 & 0.87 \\
\bottomrule
\end{tabular}
\end{table}

\begin{figure*}
    \centering
    \includegraphics[width=2\columnwidth]{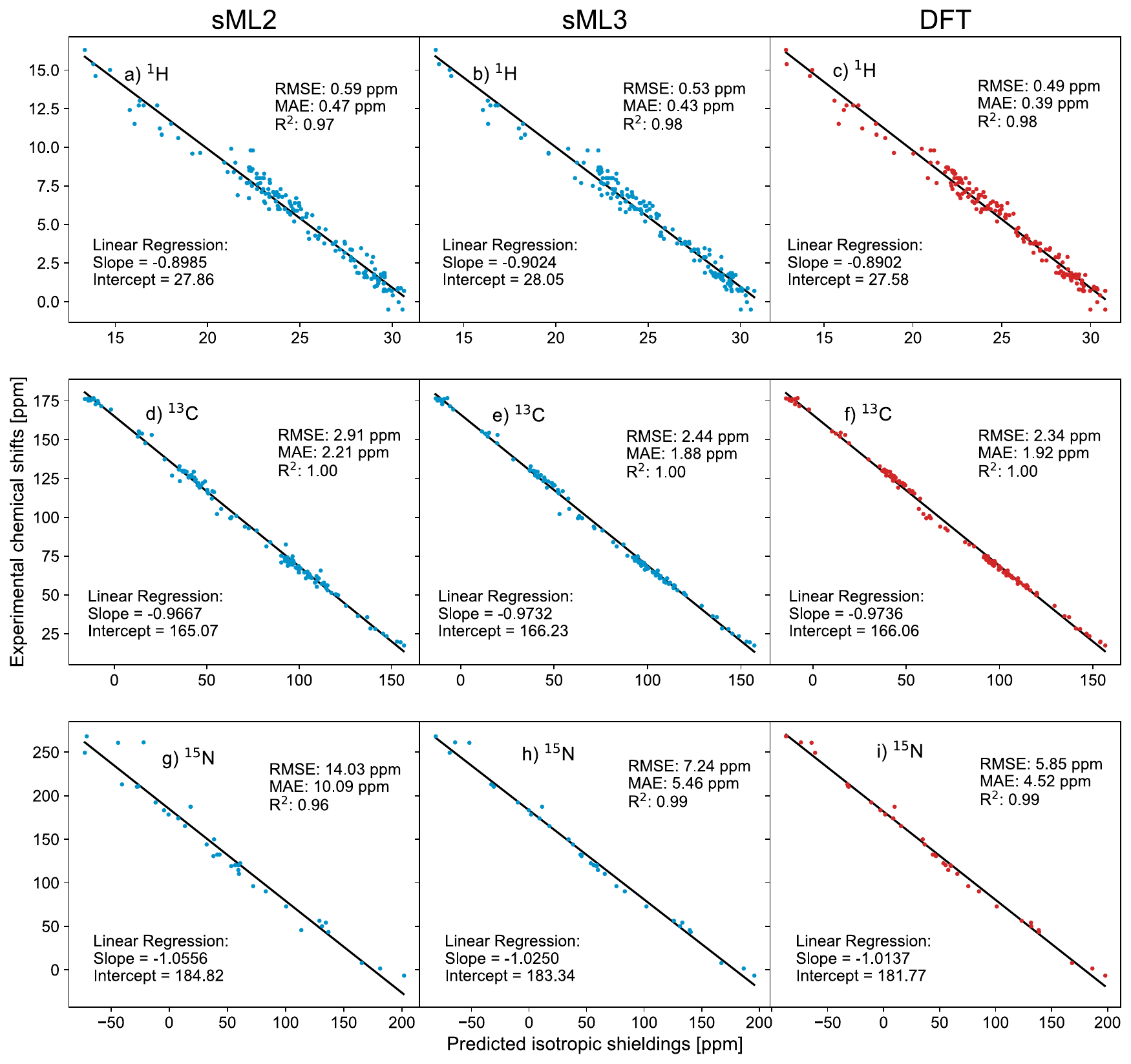}
    \caption{Parity plots between experimental chemical shifts and
predictions from \shiftmltwo{}, \shiftmlthree{} and DFT calculations for (a,b,c) $^1$H, (d, e, f) $^{13}$C,
and (g, h, i)$^{15}$N evaluated on the benchmark structures from Ramos \textit{et al.}~\cite{ramosInterplayDensityFunctional2024} for $^{13}$C and $^{15}$N, and the 13 structures where \shiftmltwo{} was evaluated for $^1$H. 
The black line shows the linear regression between experimental chemical shifts and predicted isotropic shieldings with their slope and intercept indicated in each respective subplot.}
    \label{fig:sML_vs_experimental}
\end{figure*}

We then evaluate the chemical shielding tensor predictions of the \shiftmlthree{} model on the \textit{CSD-test} set, and report the prediction accuracies of symmetric tensor components in Table~\ref{tab:cs_tensor_derived_values}. 
More information, including parity plots for $^{13}$C and $^{15}$N anisotropic shieldings can be found in section~\ref{sec: parity_plots_PAS} of the supporting information.  We report prediction accuracies for the combined principal components $\sigma_{\text{PAS}}$ (principal axis system) and combined tensor components $\sigma_{ij}$.
We report averaged prediction errors over all tensor components or principal components and across atomic environments in the test dataset, consistent with the error metric reported for chemical shieldings, which themselves are the average of the three principal components of the chemical shielding tensor. We obtain averaged eigenvalues of the ensemble by first computing the eigen decomposition of the individual committee member predictions and then averaging over the committee predictions of principal components.

\begin{figure*}
    \centering
    \includegraphics[width=1.4\columnwidth]{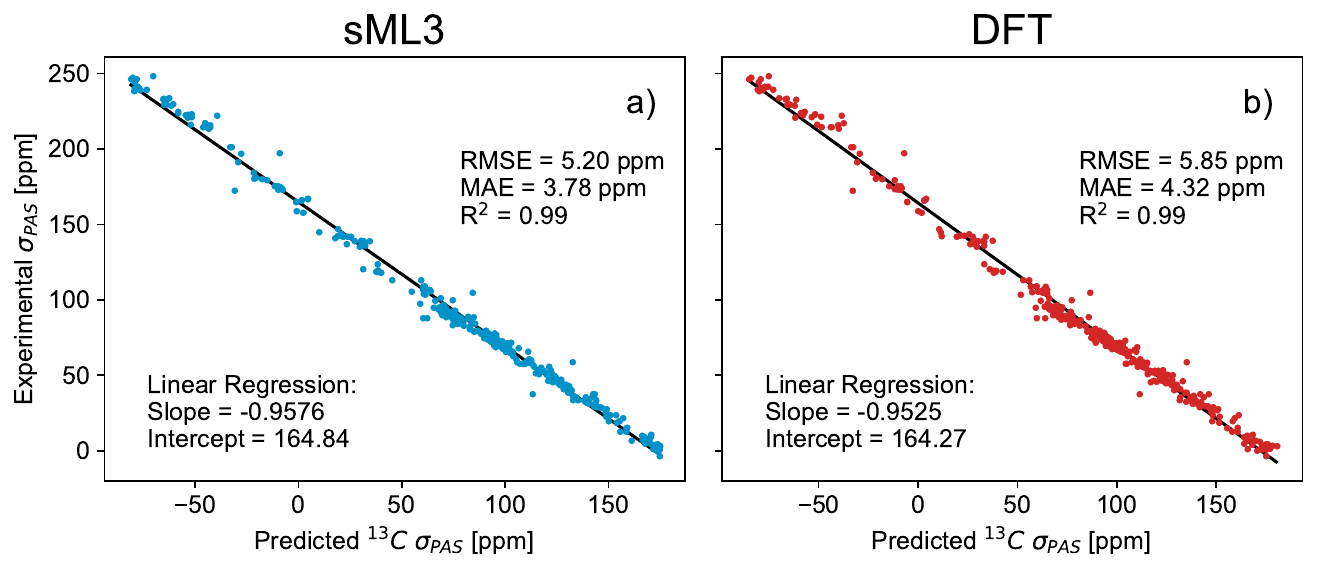}
    \caption{Parity plots between experimental principal components $\sigma_{PAS}$ of the CSA tensor and predicted components using \shiftmlthree{} for the CSA components evaluated on a subset of  benchmark structures from Ramos~\textit{et~al.}~\cite{ramosInterplayDensityFunctional2024} The black line shows the linear regression between experimental and predicted CSA principal components, with their slope and intercept shown in each respective subplot.}
    \label{fig:sML_vs_experimental_PCA}
\end{figure*}

The improvements in prediction accuracy against the DFT reference also translate to an increased accuracy of ShiftML predictions against experiments. 
We find that on a representative benchmark set of experimentally measured $^1$H and $^{13}$C shieldings,~\cite{cordova_machine_2022, ramosInterplayDensityFunctional2024} \shiftmlthree{} predicts chemical shieldings with accuracies almost on par with GIPAW DFT, namely: 0.53(0.49)~ppm for $^1$H and 2.4(2.3)~ppm for $^{13}$C, for \shiftmlthree{}(DFT). 
\rev{Given that the errors of \shiftmlthree{} against DFT are of similar size, this hints at substantial error cancellation. We also note that both computational predictions are affected by non-negligible error contributions due to the use of a static structure, neglecting finite-temperature effects and quantum fluctuations~\cite{enge+21jpcl}.}
Figure \ref{fig:sML_vs_experimental} shows parity plots of isotropic chemical shieldings estimated with \shiftmltwo{}, \shiftmlthree{} and DFT with respect to experimental chemical shifts.
DFT values were evaluated by recomputing the NMR shieldings using GIPAW~\cite{pickardAllelectronMagneticResponse2001a} on the Quantum ESPRESSO software~\cite{QE-2009, QE-2017} with the same DFT parameters as in  Ref.~\citenum{cordova_machine_2022}. \rev{Quantum ESPRESSO GIPAW calculations for these 48 structures took approximately 80 hours on 32 cores of a modern compute node for all three benchmark sets. In contrast, ShiftML3 yielded shielding predictions for all structures in just 41 seconds on a personal computer (MacBook Pro M1), demonstrating speed-ups of over three orders of magnitude.}
The conversion between experimental chemical shifts and predicted shieldings is performed by fitting globally the shieldings across all test compounds to a linear form, also shown in figure~\ref{fig:sML_vs_experimental}. Further details about the experimental benchmark are discussed in section~\ref{sec:SI_experimental_benchmark_set} of the supporting information.

We also assess the accuracy of the CSA predictions against experiments, based on a benchmark of $^{13}$C shieldings for a set of structures whose experimental shifts had been reported in Ref.~\citenum{hartmanFragmentbased13CNuclear2015}. 
The accuracy in terms of the three principal components ($\sigma_{11}, \sigma_{22}, \sigma_{33}$) of the shielding tensor is on par with that of DFT. The RMSE for \shiftmlthree{} (5.2~ppm) is even smaller than for DFT (5.8~ppm) which is likely fortuitous (Fig. \ref{fig:sML_vs_experimental_PCA}).

Overall, we find that the improved accuracy of \shiftmlthree{} against DFT-computed isotropic shieldings, relative to \shiftmltwo{}, is reflected in a corresponding improvement of the match with experiments, that is now essentially limited by the accuracy of the reference DFT calculations. 
The tensorial predictions  also have a DFT-level match with experiments, providing a further diagnostic to support structure determination and provide mechanistic insights.

\grayout{\section{Discussion}
Here we present a new machine learning model which improves upon the accuracy of previous models by up to two-fold. This improvement in resolving power goes beyond \textsuperscript{1}H and is now extended to heavy atoms, notably for \textsuperscript{13}C and \textsuperscript{15}N atomic sites. Additionally, this model now includes the ability to predict chemical shift anisotropy for organic molecular solids.

Furthermore, evaluation of predicted chemical shifts with respect to experimental chemical shifts shows an improvement in $^1$H, $^{13}$C and $^{15}$N. In Table \ref{tab:sML_vs_exp},  Improvement of 15\%, 25\% and 50\% are observed for chemical shift predictions of $^1$H, $^{13}$C and $^{15}$N predictions with respect to the previous \shiftmltwo{}.  Hence, considering that the same training set of structures as in \shiftmltwo{} has been used in \shiftmlthree{}, an improvement in accuracy must be remarked.}

To summarize our findings, we have presented a machine learning model for the prediction of chemical shieldings in organic crystals, using an unconstrained graph neural network architecture that delivers chemical shielding tensors and uncertainty quantification through an ensemble structure. 
We have shown that across all nuclei that are well-represented in the dataset, \shiftmlthree{} predicts accurate shielding tensors and isotropic shieldings in organic crystals. The isotropic prediction RMSE of 0.43~ppm for $^1$H  and 2.3~ppm for $^{13}$C against GIPAW reference calculations are a major improvement over previous ShiftML versions based on symmetry-adapted kernel models. 
The improved prediction accuracies translate to an accuracy against experimental benchmarks that practically matches that of DFT calculations, improving the confidence of NMR-guided structural determination. 
The availability of chemical shielding anisotropy, with an accuracy that also approaches that of DFT, facilitates comparison with other types of NMR experiments, providing structural and mechanistic insights. The easy-to-use, fast, and openly available implementation will facilitate usage for a broad community interested in the theoretical and experimental study of organic materials by means of NMR.

\begin{acknowledgments}
\subsection{ACKNOWLEDGEMENTS}

We thank Filippo Bigi and Michelangelo Domina for assistance with their nanoPET-IST implementation and Sergey Pozdnyakov for adapting the original PET architecture for isotropic shieldings. We thank Joseph William Abbott for the provision of a $\lambda$-SOAP model, which we used to prototype chemical shielding anisotropy predictions. This work has been supported by Swiss National Science Foundation Grant No. 200020\_212046, and by the NCCR MARVEL, funded by the Swiss National Science Foundation (grant number 182892). MC also acknowledged funding from the ERC Horizon 2020 Grant No. 101001890-FIAMMA.

\subsection{SUPPORTING INFORMATION }

The supporting information is provided below,
containing additional details on dataset and model construction, uncertainty quantification, and the usage of the \shiftmlthree{} Python package, which is available under an open-source license at \url{https://github.com/lab-cosmo/shiftml}.
The training, validation and testing datasets containing chemical shielding tensor data as well as training input files, model files and evaluation scripts will be made available upon publication via the public data record associated with this publication.~\cite{kellner2025deep}

\end{acknowledgments}

\section{References}
\clearpage

\onecolumngrid
\part*{Supplementary Information}

\setcounter{secnumdepth}{3}           %
\renewcommand{\thesection}{S\arabic{section}}
\setcounter{section}{0}

\renewcommand{\thefigure}{S\arabic{figure}}
\setcounter{figure}{0}

\renewcommand{\thetable}{S\arabic{table}}
\setcounter{table}{0}

\section{Detailed prediction accuracies of \shiftmlthree} \label{sec:detailed_preds_sML5}

We evaluate the accuracy of \shiftmlthree{} on the CSD-test set described in the main text. In Table~\ref{tab:complete_metrcis}, we report the size of the training, validation and CSD-test set together with the mean absolute error (MAE), root-mean-squared-error (RMSE), the coefficient of determination (R\textsuperscript{2}), and the standardized RMSE (\%) evaluation metrics of $\sigma_{\text{iso}}$ predictions of \shiftmlthree{} for all chemical species contained in CSD-test.

\begin{table}[htbp]
  \centering
  \caption{Predicted accuracy of $\sigma_{\text{iso}}$ of \shiftmltwo{} (\shiftmltwoshort{}) and \shiftmlthree{} (\shiftmlthreeshort{}) on the CSD-test hold-out set}
    \resizebox{\textwidth}{!}{\begin{tabular}{c|@{\hskip 2mm}c@{\hskip 2mm}c@{\hskip 2mm}c@{\hskip 2mm}|@{\hskip 6mm}cc@{\hskip 6mm}cc@{\hskip 6mm}cc@{\hskip 6mm}cc}
    \toprule
    Nucleus & \hspace{4pt} N$_{\text{train}}$\hspace{4pt}  &  N$_{\text{validation}}$  & \hspace{4pt}N$_{\text{test}}$ \hspace{4pt} & \multicolumn{2}{c@{\hskip 6mm}}{MAE [ppm]} & \multicolumn{2}{c@{\hskip 6mm}}{RMSE [ppm]} & \multicolumn{2}{c@{\hskip 6mm}}{RMSE [\%]} & \multicolumn{2}{c}{R$^2$} \\
    \hline
          &       &       &       & \shiftmltwoshort{}  & \shiftmlthreeshort{}  & \shiftmltwoshort{}  & \shiftmlthreeshort{}  & \shiftmltwoshort{}  & \shiftmlthreeshort{} & \shiftmltwoshort{}  & \shiftmlthreeshort{} \\
\hline
    $^{1}$H & 547,887& 18,505 & 47,706 & 0.39  & 0.33  & 0.51  & 0.43  & 16    & 14    & 0.97  & 0.98 \\
    $^{13}$C & 437,529& 16,971 & 43,435 & 3.15  & 1.58  & 4.63  & 2.32  & 9     & 5     & 0.99  & 1.00 \\
    $^{15}$N & 119,413& 1,879  & 4,635  & 10.29 & 5.00  & 15.72 & 10.41 & 14    & 10    & 0.98  & 0.99 \\
    $^{17}$O & 135,305& 3,272  & 8,058  & 16.11 & 7.51  & 22.96 & 11.45 & 13    & 6     & 0.98  & 1.00 \\
    $^{19}$F & 37,765& 296   & 569   & 7.77  & 4.42  & 10.83 & 6.66  & 18    & 11    & 0.97  & 0.99 \\
    $^{33}$S & 30,500& 386   & 1,084  & 34.35 & 15.05 & 53.34 & 26.34 & 35    & 17    & 0.88  & 0.97 \\
    $^{31}$P & 8,508& 84    & 151   & 21.83 & 9.49  & 39.20 & 18.98 & 61    & 30    & 0.62  & 0.91 \\
    $^{35}$Cl & 25,608& 188   & 569   & 17.40 & 10.44 & 23.85 & 15.11 & 21    & 13    & 0.96  & 0.98 \\
    $^{23}$Na & 1,138& 9     & 5     & 7.64  & 4.09  & 8.10  & 4.61  & 100   & 57    & 0.00  & 0.68 \\
    $^{43}$Ca & 613& 3     & 5     & 5.81  & 2.96  & 6.98  & 3.34  & 44    & 21    & 0.81  & 0.96 \\
    $^{25}$Mg & 283& 3     & 7     & 10.69 & 3.68  & 14.55 & 5.49  & 58    & 22    & 0.66  & 0.95 \\
    $^{39}$K & 995& 3     & 6     & 3.42  & 4.29  & 4.28  & 4.65  & 48    & 52    & 0.77  & 0.73 \\
    \bottomrule
    \hline
    \end{tabular}%
    }
  \label{tab:complete_metrcis}%
\end{table}%

\section{Rotational fluctuations of PET and PET-ensemble} \label{sec:SI_rotationa_fluc}
PET models learn the correct rotational behaviour of the targets via rotational augmentations. 
Langer and coworkers have investigated the effect of relaxing exact rotational invariance to learned invariances,~\cite{langerProbingEffectsBroken2024} and found that for realistic atomistic modelling tasks, this constraint can be safely relaxed. 
The remaining rotational fluctuations of the PET model and PET ensemble should however be carefully assessed for chemical shielding prediction tasks. Unlike the prediction of potential energies, or other structure-wise global quantities, chemical shieldings are atom-wise properties.
The common ansatz of predicting potential energies with neural networks, is by summing over local, per-atom terms. Shortcomings (imperfectly learned rotational invariance) of a non-invariant architecture might simply be compensated by error cancellation summing over atomic contributions. Since chemical shieldings and shielding tensors are exactly these local contributions (one prediction per atom), we study the effects of applying random rotations to structures and measuring the fluctuations of the predicted isotropic chemical shieldings. We apply 30 random rotations to each structure in the CSD-test set and evaluate the standard deviation of the isotropic chemical shielding predictions. For a perfectly invariant model, these fluctuations should be exactly zero due to the invariance of the model. For an approximately invariant model, these predictions should be quite small, at most a fraction of the prediction RMSEs, such that they do not interfere with NMR crystal structure prediction workflows. In Table~\ref{tab:rot_cs_iso_fluc} we list rotational fluctuations of \shiftmlthree{} predictions across various chemical species and compare them with the rotational fluctuations of single PET models, highlighting the strongly reduced fluctuation of the ensemble of PET models that make up ShiftML3.

\begin{table}[!ht]
\caption{\label{tab:rot_cs_iso_fluc} Average rotational fluctuations (average of all environment-wise standard deviations applying 30 random rotations) std$_{\text{rot}}$[$\sigma_{\text{iso}}$] and maximal rotational fluctuation (of environment-wise standard deviations)  across all considered environments max$_{\text{rot}}$[$\sigma_{\text{iso}}$] of \shiftmlthree{} predictions across the CSD-test set, of both the final \textit{ensemble} model, as well as the individual committee models (single). All metrics are given in (ppm). \textit{Single} is obtained by first evaluating each model on the hold-out test set and then averaging over the mean and max fluctuations of each model on the entire hold-out set,  whilst the \textit{ensemble} values are obtained by first averaging the individual model predictions (computing the \shiftmlthree{} model predictions) and then evaluating mean and max fluctuations on the hold-out set.}
\begin{tabular}{c|cc|cc}
\toprule
Nucleus & \multicolumn{2}{c|}{single} & \multicolumn{2}{c}{ensemble (\shiftmlthree{})} \\
\hline
 & std$_{\text{rot}}$[$\sigma_{\text{iso}}$] & max$_{\text{rot}}$[$\sigma_{\text{iso}}$] & std$_{\text{rot}}$[$\sigma_{\text{iso}}$] & max$_{\text{rot}}$[$\sigma_{\text{iso}}$] \\
\hline
$^{1}\text{H}$ & 0.41 & 1.78 & 0.06 & 0.23 \\
$^{13}\text{C}$ & 1.19 & 7.13 & 0.14 & 0.71 \\
$^{15}\text{N}$ & 2.57 & 23.49 & 0.29 & 1.57 \\
$^{17}\text{O}$ & 3.98 & 43.35 & 0.45 & 2.22 \\
$^{19}\text{F}$ & 2.27 & 9.09 & 0.26 & 0.70 \\
$^{33}\text{S}$ & 5.28 & 36.90 & 0.59 & 3.13 \\
$^{31}\text{P}$ & 2.82 & 19.81 & 0.33 & 1.49 \\
$^{35}\text{Cl}$ & 4.73 & 22.13 & 0.50 & 2.03 \\
\bottomrule
\end{tabular}
\end{table}

We find that for most nuclei, ensemble uncertainty estimates are correlated with the rotational fluctuations.
\begin{figure}
    \centering
    \includegraphics[width=0.6\linewidth]{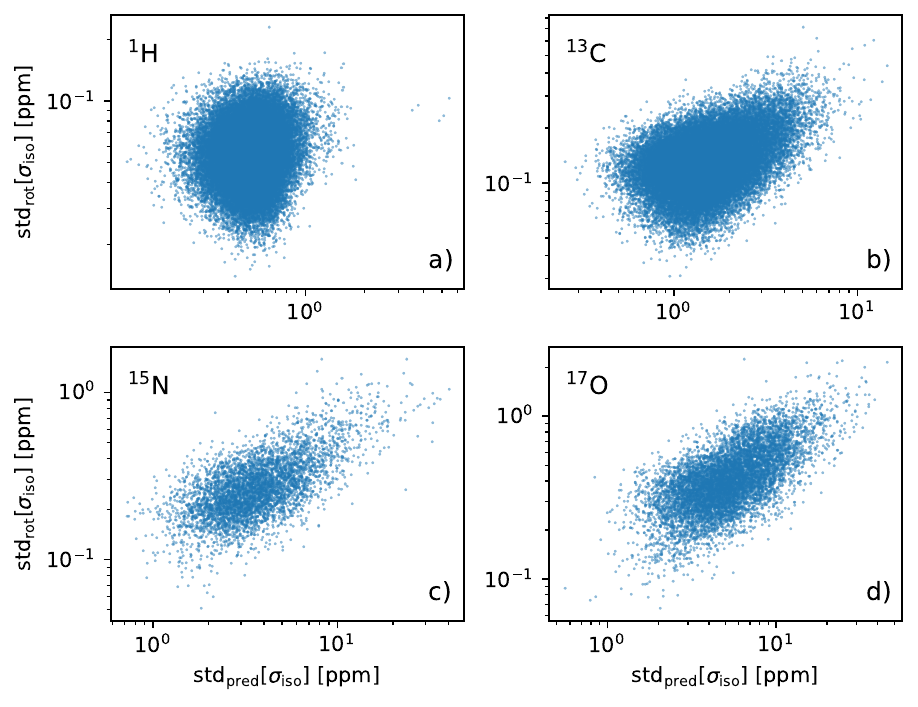}
        \caption{Correlations of uncertainty estimates $\text{std}_{\text{pred}}[\sigma_{\text{iso}}]$ across all atomic environments in the CSD-test set with the rotational fluctuations of the same environments std$_{\text{rot}}[\sigma_{\text{iso}}]$, for $^1$H (a), $^{13}$C (b), $^{15}$N (c) and $^{17}$O (d).}

    \label{fig:XX}
\end{figure}
We evaluate the structure-wise fluctuations of the RMSEs of the \shiftmlthree{} predictions for CSD-test with respect to the GIPAW reference. To assess the effect of approximate invariance we also compute the fluctuations of the RMSE between all $^1$H environments in a structure and its non-rotated structure. The mean fluctuation (mean of the standard deviation of fluctuations) of this RMSE across the CSD-test set is 0.0065 ppm and the maximal fluctuation we find to be 0.0271 ppm.

\section{Model evaluation timings}
\label{sec:SI_computational_timing}

\begin{figure}
    \centering
    \caption{Scaling of the inference time of the \shiftmlthree{} model with system size. Blue lines indicate timings for diamond supercells of increasing size. Black markers are the evaluation times of the 1PGA protein crystal. Grey lines correspond to linear scaling. }
    \includegraphics[width=0.5\linewidth]{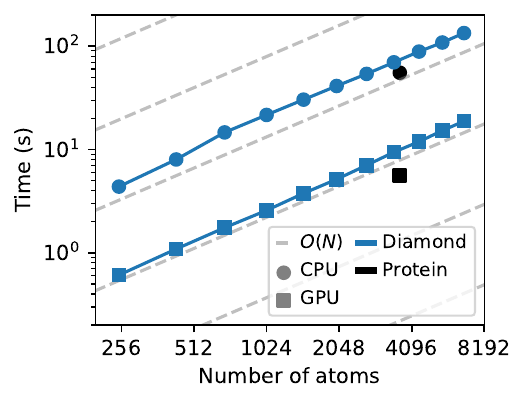}

    \label{fig:scaling_speed}
\end{figure}

We benchmark the evaluation time of the \shiftmlthree{} model. All benchmark timings are obtained via the ASE interface of the ShiftML library. We report timings that include setting up the ASE calculator once and then computing chemical shielding tensors for a series of structures. We believe that this protocol reflects most accurately the evaluation timings for a common NMR crystallography workflow. The timings via the ASE interface include also the neighbour list construction.
The ASE interface does not support batched evaluations, or efficient parallelization over ensemble models for smaller system sizes. 
We report CPU-only and CPU+GPU timings. We report evaluation times on a Lenovo workstation with an AMD Ryzen Threadripper PRO 3955WX CPU with 16 cores. CPU+GPU timings are obtained on the same workstation, using an NVIDIA GeForce RTX 4070 Ti SUPER consumer GPU, which is representative of a configuration that does not require access to high-end AI-dedicated hardware.
As shown in Table~\ref{tab:timings}, even a low-end GPU accelerates by 3x-4x the evaluation of structures with the size of typical small-molecule pharmaceuticals. 
The acceleration is also evident at larger system sizes: the scaling of the \shiftmlthree{} model for 
diamond supercells of increasing size is almost perfectly linear, and there is an order-of-magnitude acceleration when using a GPU (Fig.~\ref{fig:scaling_speed}).
These timings are also representative of large biomolecules, as hinted at by the inference timings for the protein crystal of 1PGA~\cite{gallagherTwoCrystalStructures1994}, the B1 immunoglobulin-binding domain of the streptococcal protein G with 56 amino acid residues and 3,656 atoms in the conventional cell (also shown in Figure~\ref{fig:scaling_speed}). 1PGA was a target of several solid-state NMR studies~\cite{wylieSiteSpecific13CChemical2005, franksMagicAngleSpinningSolidState2005} and therefore lies well within the range of atomistic systems of interest for which chemical shielding predictions may complement experimental studies. A computation of chemical shieldings for this protein takes 5.6 s using the GPU. Moving to even larger systems, we find that on the consumer GPU used for benchmarking with 16 GB of VRAM, we can predict $^{13}$C shieldings in diamond supercells of up to 16,000 atoms before exhausting the graphics card video memory, demonstrating the memory efficiency of \shiftmlthree{}.

\begin{table}
    \centering
\caption{Average timing of the \shiftmlthree{} model for evaluating chemical shieldings on CPU or GPU. We also list the average number of atoms $\overline{\text{N}}_{\text{atoms}}$ in the CSD-test hold-out test set. }
\label{tab:timings}
    \begin{tabular}{c c c} \toprule  
         Set& \makecell{Time per struct. [s]\vspace{0.5mm} \\  CPU/GPU} &  $\overline{\text{N}}_{\text{atoms}}$ \\ \midrule  
         CSD-test&  0.81/0.23& 125\\    \bottomrule 
    \end{tabular}
\end{table}

\section{Implementation} \label{sec:shiftml}
\label{SI:implementation}
To simplify usage of \shiftmlthree{} for inference, we make it available with an easy-to-install Python package, released together with this publication. ShiftML is available from the Python Package Index (PyPI) and can be easily installed with

\begin{verbatim} pip install shiftml \end{verbatim}

The \shiftmlthree{} model integrates into existing materials modelling and crystal structure prediction workflows via the Atomic Simulation Environment (ASE). Chemical shielding predictions can be obtained by initializing an ASE calculator object and reading in the geometry file in one of the various supported file formats. 
Once the calculator is initialized, the model weights are automatically downloaded and loaded by the model.
A minimalistic usage example is given below, and a more thorough tutorial is  available at \url{http://atomistic-cookbook.org/shiftml/shiftml-example.html}.

\begin{verbatim}
from ase.build import bulk
from shiftml.ase import ShiftML

frame = bulk("C", "diamond", a=3.566)
model = ShiftML("ShiftML3")

Ypred = model.get_cs_tensor(frame)
\end{verbatim}

\section{Uncertainty quantification and calibration}
\label{sec:SI_UQ}
Ensemble uncertainty estimates are typically miscalibrated (as is often the case~\cite{guoCalibrationModernNeural2017a, kuleshovAccurateUncertaintiesDeep2018c} in uncertainty quantification schemes for neural networks), i.e. the 
uncertainty estimates across all samples of a hold-out validation or test set model uncertainty estimates are systematically over- or underconfident. 
It is possible to rescale the uncertainty estimates, and correct the miscalibration by computing a global, input-independent scaling factor~\cite{musi+19jctc,imbalzanoUncertaintyEstimationMolecular2021} (an empirical correction that is similar in spirit to \emph{temperature scaling},~\cite{guoCalibrationModernNeural2017a} a classical method in uncertainty quantification). 
The scaling factor can be computed via a closed-form expression, that can be derived by maximizing the Gaussian negative log-likelihood on a hold-out test set given predicted uncertainties $\sigma^{2}_{\text{pred}}(A_i)$ and squared prediction errors $z^2(A_i)$ of some environment $A_i$.
\begin{equation}
    \alpha^2 = \frac{1}{N_{\text{samples}}}\sum^{N_{\text{samples}}}_{i=1} \frac{z^2(A_i)}{\sigma^{2}_{\text{pred}}(A_i)}
\end{equation}
The final calibrated uncertainties are then obtained as $\alpha^2\sigma^2_\text{pred}(A_i)$ - although in practice we scale the ensemble members around their mean, which simplifies uncertainty propagation.~\cite{kellnerUncertaintyQuantificationDirect2024a} 
We find that for isotropic chemical shieldings and shielding tensor principal components the calibration factors $\alpha$ are moderate and always close to $1$, indicating that the \shiftmlthree{} ensemble uncertainty estimates are already well calibrated.

\begin{figure}
    \centering
    \includegraphics[width=0.6\linewidth]{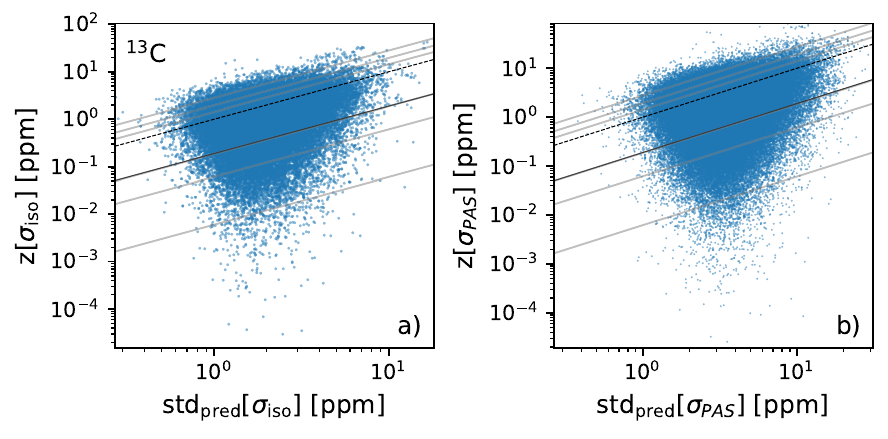}
    \caption{Predicted uncertainties of $^{13}$C isotropic chemical shieldings $\text{std}_{\text{pred}}[\sigma_{\text{iso}}]$ vs the actual prediction errors $\text{z}[\sigma_{\text{iso}}]$ (a) and predicted uncertainties of the corresponding principal components $\text{std}_{\text{pred}}[\sigma_{\text{PAS}}]$ vs the actual component prediction errors $\text{z}[\sigma_{\text{PAS}}]$ (b). The grey lines indicate quantiles of the expected error distribution for a given predicted uncertainty value and are intended to guide the reader's eye: For example, on the test set, for well calibrated uncertainty estimates there should be 90\% of the samples between the two outermost lines as they correspond to quantile lines of 95\% and 5\% of the log-folded normal distribution. }
    \label{fig:SI-uq-parity-plot}
\end{figure}

We evaluate the uncertainty estimates of the \shiftmlthree{} model on the CSD-test set and plot the predicted uncertainties of $^{13}$C shieldings vs the actual prediction errors. In Figure~\ref{fig:SI-uq-parity-plot} we show parity plots of predicted and calibrated uncertainties of the \shiftmlthree{} model and the actual errors on a hold-out test set.

\section{Antisymmetric chemical shielding tensor predictions} 
\label{sec:antisymm_preds}
\begin{figure}[!ht]
    \centering
    \includegraphics[width=0.5\linewidth]{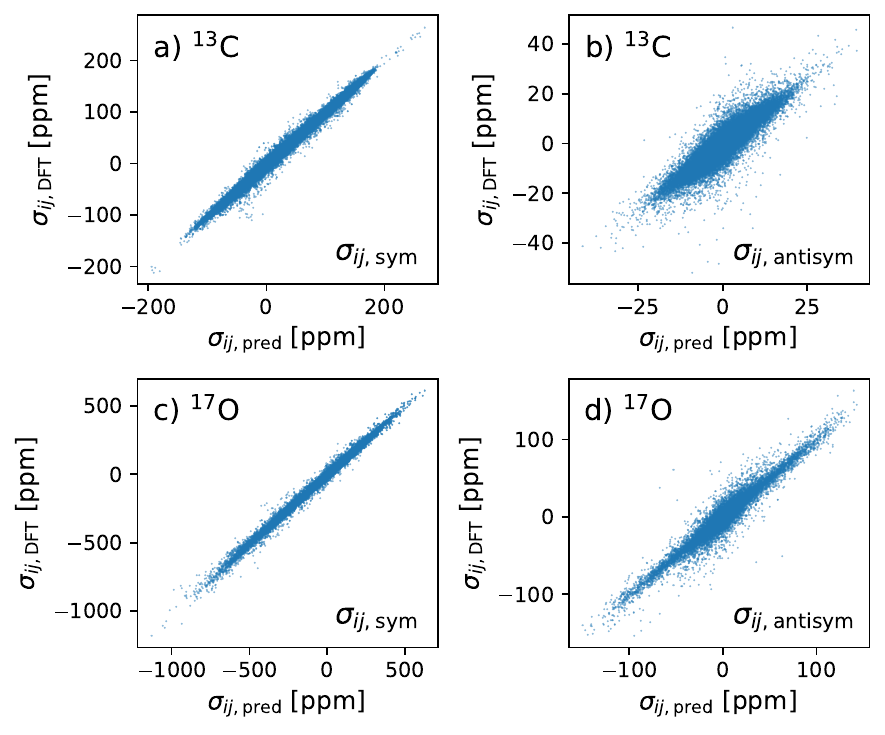}
    \caption{Parity plots of symmetric and antisymmetric chemical shielding tensor components $\sigma_{ij,\text{sym}}$,
    $\sigma_{ij,\text{antisym}}$ for $^{13}$C (a,b) and $^{17}$O (c,d) comparing the values predicted with \shiftmlthree{} and those computed with DFT.}
    
    \label{fig:sym_v_antisym}
\end{figure}

The chemical shielding tensor can be decomposed into a symmetric and antisymmetric part (eq.~\ref{eq:tensor_decomposition}):
\begin{equation}
    \label{eq:tensor_decomposition}
    \sigma = \sigma_{\text{sym}} + \sigma_{\text{antisym}} = \frac{1}{2} (\sigma + \sigma^\text{T}) + \frac{1}{2} (\sigma - \sigma^\text{T})
\end{equation}
Only the symmetric part enters directly into the NMR spectrum, whilst the antisymmetric part affects relaxation times.~\cite{mcconnellAnisotropicChemicalShielding1956a} Therefore, in the main part of this manuscript we have focussed on evaluating and utilizing the CSA, $\sigma_{\text{PAS}}$ which are the principal components of $\sigma_{\text{sym}}$. \shiftmlthree{} can predict the full chemical shielding tensor, containing symmetric and antisymmetric parts. In Figure~\ref{fig:sym_v_antisym} we show parity plots of the symmetric and antisymmetric components $\sigma_{ij,\text{sym}}$ and $\sigma_{ij,\text{antisym}}$  of the $^{13}$C and $^{17}$O chemical shielding tensors.

In Table~\ref{tab:metrics_anti_sym_sym} we list the prediction accuracies of \shiftmlthree{} on CSD-test separately on antisymmetric and symmetric components of the chemical shielding tensor.
\begin{table}[!ht]
    \caption{Prediction accuracies RMSEs (ppm) and normalized prediction accuracies RMSE/std (\%) of \shiftmlthree{} of the components of the symmetric $\sigma_{ij,\text{sym}}$ and antisymmetric shielding tensor $\sigma_{ij,\text{antisym}}$. }
\begin{tabular}{c|@{\hskip 4mm}cc@{\hskip 4mm}cc@{\hskip 4mm}}
\toprule
Nucleus & \multicolumn{2}{c@{\hskip 4mm}}{$\sigma_{ij,\text{sym}}$} & \multicolumn{2}{c@{\hskip 4mm}}{$\sigma_{ij,\text{antisym}}$} \\
\hline
 & [ppm] & [$\%$] & [ppm] & [$\%$] \\
\hline
$^{1}\text{H}$ & 0.72 & 6 & 0.52 & 57 \\
$^{13}\text{C}$ & 3.08 & 5 & 2.28 & 41 \\
$^{15}\text{N}$ & 10.75 & 11 & 5.98 & 47 \\
$^{17}\text{O}$ & 12.09 & 7 & 7.82 & 27 \\
$^{19}\text{F}$ & 6.28 & 5 & 3.35 & 39 \\
$^{33}\text{S}$ & 30.37 & 18 & 18.10 & 52 \\
$^{31}\text{P}$ & 18.73 & 14 & 11.02 & 72 \\
$^{35}\text{Cl}$ & 15.32 & 5 & 9.57 & 47 \\
$^{23}\text{Na}$ & 4.01 & 2 & 1.58 & 81 \\
$^{43}\text{Ca}$ & 3.85 & 1 & 1.31 & 29 \\
$^{25}\text{Mg}$ & 9.79 & 3 & 4.55 & 39 \\
$^{39}\text{K}$ & 3.95 & 1 & 2.05 & 67 \\
\bottomrule
\end{tabular}
    \label{tab:metrics_anti_sym_sym}
\end{table}

\begin{figure}
    \centering
    \includegraphics[width=0.5\linewidth]{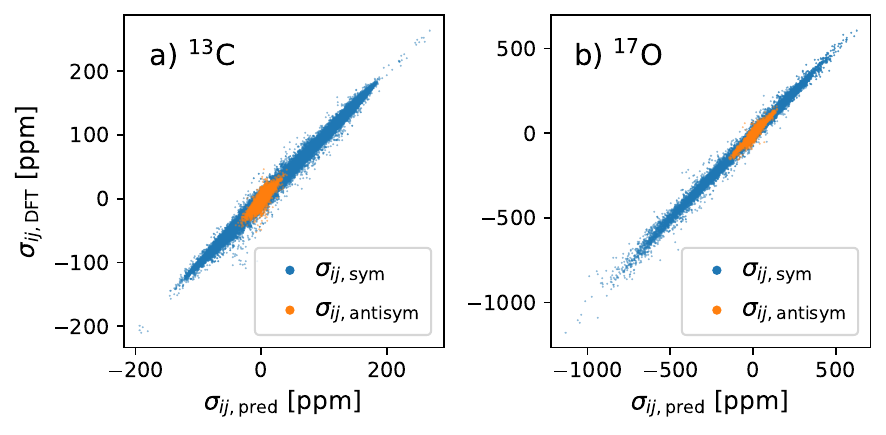}
    \caption{Parity plots of symmetric and antisymmetric chemical shielding tensor components $\sigma_{ij,\text{sym}}$,
    $\sigma_{ij,\text{antisym}}$ for $^{13}$C (a) and $^{17}$O (b) - combined parity plots.}

    \label{fig:sym_v_antisym_combined}
\end{figure}

Figure~\ref{fig:sym_v_antisym_combined} shows that the symmetric and antisymmetric tensor components (for nuclei like $^{13}$C and $^{17}$O) have similar absolute prediction errors. However, because the antisymmetric values span a much smaller range, their relative errors are much larger than those of the symmetric components.

\label{sec: parity_plots_PAS}

\begin{figure}[!ht]
    \centering
    \includegraphics[width=0.6\linewidth]{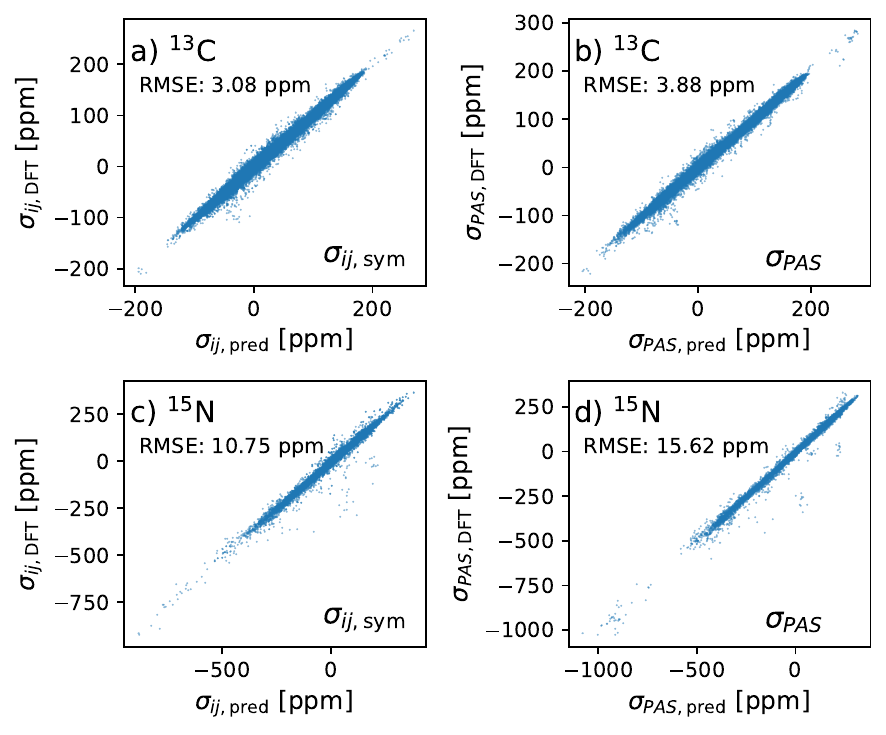}
    \caption{Parity plots of principal tensor components $\sigma_{\text{PAS}}$ and symmetric tensor components $\sigma_{ij}$ for $^{13}$C (a,b) and $^{15}$N (c,d)}
    \label{fig:SI_parity_plot_PAS}
\end{figure}

In Figure~\ref{fig:SI_parity_plot_PAS} we show parity plots of symmetric shielding tensor component predictions and principal tensor component predictions.

\begin{figure}[!ht]
    \centering
    \includegraphics[width=0.8\linewidth]{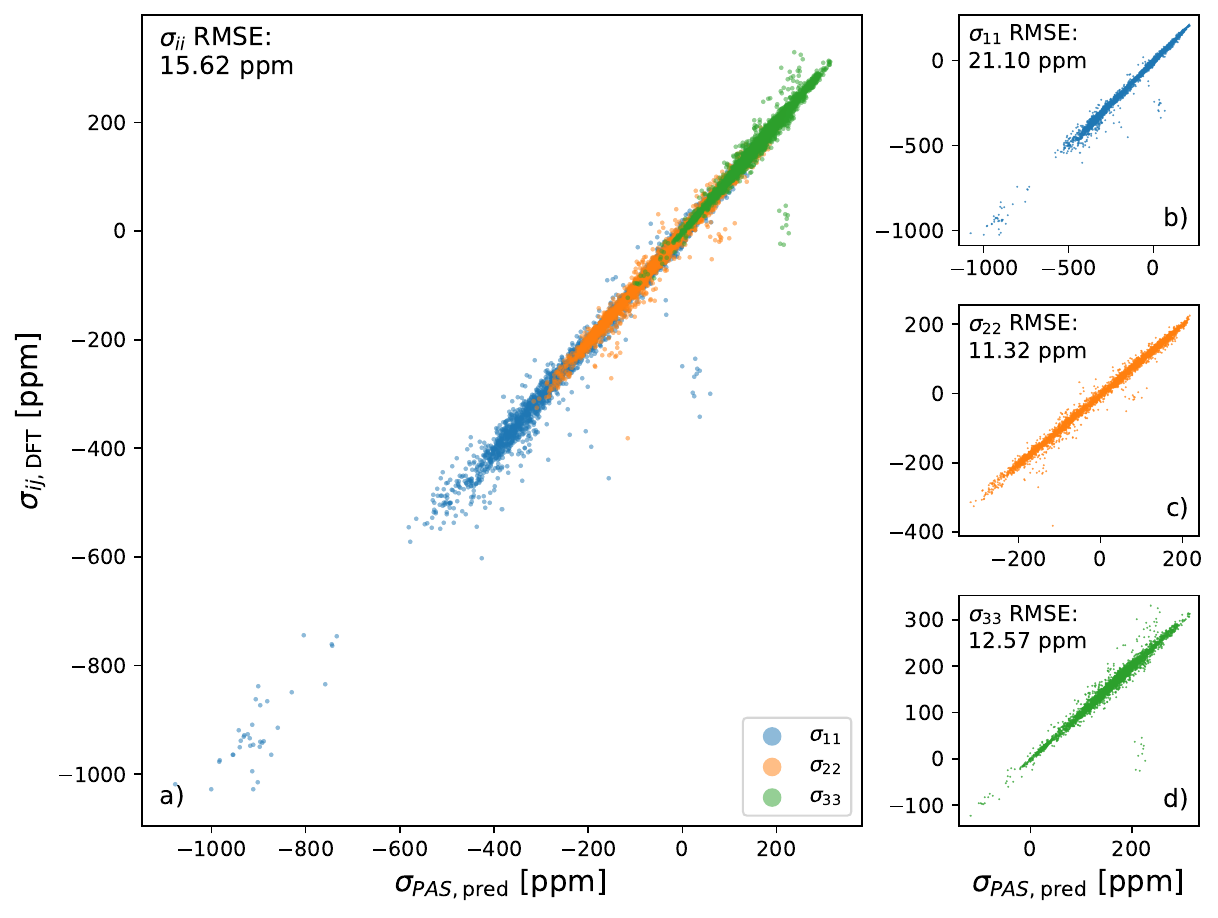}
    \caption{\rev{Component resolved parity plots of the principal tensor components of $^{15}$N shielding tensors. Combined principal components are plotted in a), $\sigma_{\text{11}}$ in b), $\sigma_{\text{22}}$ in c) and $\sigma_{\text{33}}$ in d).}}
    \label{fig:SI_parity_plot_PAS_resolved}
\end{figure}

\rev{In Figure~\ref{fig:SI_parity_plot_PAS_resolved} we show parity plots of the principal tensor components resolved for individual principal components $\sigma_{11}$,$\sigma_{22}$ and $\sigma_{33}$. In Table~\ref{tab:metrics_components_resolved} we list prediction accuracies of the individual principal components.}

\begin{table}[htbp]
  \centering
  \caption{\rev{Predicted accuracy of $\sigma_{\text{PAS}}$ of \shiftmlthreeshort{} on the CSD-test hold-out set, resolved for individual principal components $\sigma_{\text{11}}$, $\sigma_{\text{22}}$ and $\sigma_{\text{33}}$. } }
    \resizebox{0.75\textwidth}{!}{
\begin{tabular}{c|@{\hskip 6mm}c@{\hskip 2mm}c@{\hskip 2mm}c@{\hskip 6mm}|@{\hskip 6mm}c@{\hskip 2mm}c@{\hskip 2mm}c@{\hskip 6mm}|@{\hskip 6mm}c@{\hskip 2mm}c@{\hskip 2mm}c@{\hskip 6mm}}
\hline
Nucleus & \multicolumn{3}{c@{\hskip 12mm}}{MAE [ppm]} & \multicolumn{3}{c@{\hskip 12mm}}{RMSE [ppm]} & \multicolumn{3}{c@{\hskip 6mm}}{RMSE [\%]} \\
    \hline
 & $\sigma_{11}$ & $\sigma_{22}$ & $\sigma_{33}$ & $\sigma_{11}$ & $\sigma_{22}$ & $\sigma_{33}$ & $\sigma_{11}$ & $\sigma_{22}$ & $\sigma_{33}$ \\
\hline
$^{1}\text{H}$ & 0.65 & 0.60 & 0.69 & 0.86 & 0.77 & 0.90 & 21 & 22 & 21 \\
$^{13}\text{C}$ & 2.77 & 2.96 & 2.19 & 4.17 & 4.28 & 3.07 & 5 & 8 & 10 \\
$^{15}\text{N}$ & 9.55 & 6.42 & 5.66 & 21.10 & 11.32 & 12.57 & 11 & 9 & 19 \\
$^{17}\text{O}$ & 14.40 & 9.35 & 7.09 & 23.70 & 13.28 & 10.41 & 8 & 6 & 13 \\
$^{19}\text{F}$ & 6.02 & 6.35 & 4.83 & 8.99 & 9.40 & 7.45 & 11 & 10 & 20 \\
$^{33}\text{S}$ & 26.75 & 19.35 & 21.50 & 57.02 & 32.32 & 34.40 & 29 & 22 & 21 \\
$^{31}\text{P}$ & 12.87 & 11.43 & 15.75 & 23.90 & 17.49 & 28.15 & 28 & 25 & 45 \\
$^{35}\text{Cl}$ & 14.75 & 14.00 & 13.31 & 20.99 & 21.39 & 19.03 & 11 & 14 & 32 \\
$^{23}\text{Na}$ & 4.76 & 5.27 & 2.82 & 4.95 & 6.79 & 3.40 & 68 & 65 & 43 \\
$^{43}\text{Ca}$ & 4.66 & 4.98 & 3.96 & 4.74 & 5.48 & 4.52 & 23 & 30 & 34 \\
$^{25}\text{Mg}$ & 3.68 & 4.97 & 10.17 & 4.97 & 5.39 & 12.45 & 20 & 23 & 45 \\
$^{39}\text{K}$ & 5.22 & 3.31 & 4.74 & 6.12 & 3.78 & 5.43 & 79 & 42 & 50 \\
\bottomrule
\end{tabular}
}
  \label{tab:metrics_components_resolved}%
\end{table}%

\section{Details of experimental benchmark evaluations}
\label{sec:SI_experimental_benchmark_set}

\rev{
Evaluation of $^1$H was done on the same 13 structures on which \shiftmltwo{} was evaluated \cite{cordova_machine_2022}, using the same geometry-optimized structures. 
Predicted shieldings of methyl or primary amine groups were averaged to account for rapid rotation, resulting in equivalent chemical shielding for the averaged sites. Ambiguous experimental assignments were resolved by iterating over all possible permutations of the unassigned shifts—such as for methylene groups—and selecting the assignment with the lowest RMSE between predicted and experimental chemical shifts. Only the permutation with the lowest RMSE entered the global linear regression. We note that the reported RMSE comparing prediction to experiment for \shiftmltwo{} is modestly different from the original work. This difference originates from the use of a global regression compared with the on-the-fly regression used in the original work.

For $^{13}$C and $^{15}$N, the accuracies were evaluated using the benchmark of 36 structures from Ramos \textit{et al.}~\cite{ramosInterplayDensityFunctional2024}, with geometries optimized at the PBE-D3(BJ) level of theory~\cite{perdewGeneralizedGradientApproximation1996}. For the evaluation of carbon CSA parameter predictions, we used the benchmark consisting of 19 structures from Hartman \textit{et al.}~\cite{hartmanFragmentbased13CNuclear2015}.
}

\begin{table}[!ht]
  \centering

  \caption{RMSE and Linear regression parameters for converting DFT, \shiftmltwo{}, and \shiftmlthree{} prediction into chemical shifts for the respective benchmarks.  }
    \begin{tabular}{ccccc}
    \toprule
    \textbf{Nucleus} & \textbf{Method} & \textbf{RMSE} & \textbf{Slope} & \textbf{Intercept} \\
    \midrule
    $^1$H iso     & \shiftmltwo{} & 0.59  & -0.8985 & 27.86 \\
          & \shiftmlthree{} & 0.53  & -0.9024 & 28.05 \\
          & DFT   & 0.49  & -0.8902 & 27.58 \\
    \midrule
    $^{13}$C iso     & \shiftmltwo{} & 2.91  & -0.9667 & 165.07 \\
          & \shiftmlthree{} & 2.44  & -0.9732 & 166.23 \\
          & DFT   & 2.34  & -0.9736 & 166.06 \\
    \midrule
    $^{15}$N iso     & \shiftmltwo{} & 14.03 & -1.0556 & 184.82 \\
          & \shiftmlthree{} & 7.24  & -1.025 & 183.34 \\
          & DFT   & 5.85  & -1.0137 & 181.77 \\
    \midrule
    $^{13}$C PAS     & \shiftmlthree{} & 5.2   & -0.9576 & 164.84 \\
          & DFT   & 5.85  & -0.9525 & 164.27 \\
    \bottomrule
    \end{tabular}%
  \label{tab:linear_regresion_param}%
\end{table}%

\section{Comparing \shiftmlthree{} with other machine learning models for chemical shielding prediction}
A direct comparison of the \shiftmlthree{} prediction accuracies with \shiftmltwo{} reveals that it is more accurate for predicting isotropic chemical shieldings across all relevant elements. To the best of our knowledge, no other machine learning model has been trained on the latest iteration of the ShiftML datasets, including thermally distorted structures and additional chemical elements. 
Unzueta et al. trained GNN models~\cite{unzueta2022low} and Liu and coworkers a multiresolution 3D-DenseNet architecture on the CSD-2K dataset~\cite{liuMultiresolution3DDenseNetChemical2019a}, an early iteration of the ShiftML training sets. An alternative model construction strategy is the NMRNet from Xu and coworkers, that fine-tuned a transformer model for the chemical shielding prediction in liquids on the CSD-2K set~\cite{xuUnifiedBenchmarkFramework2025}. All three works report evaluation accuracies on the CSD-500 set, a set of DFT relaxed organic crystals, containing at most H,C,N and O. Unfortunately, the number of shared structures between the CSD-test and the CSD-500 test set is too small (about 10 structures) to make direct comparisons between the accuracies of \shiftmlthree{} and other neural network architectures. Shielding computations are highly sensitive to k-space discretization and choice of pseudopotentials~\cite{widdifieldNMRCrystallographyStructure2025}. CSD-2K/CSD-500 was computed with other DFT convergence parameters than the \shiftmltwo{}/\shiftmlthree{} dataset, severely hampering a direct comparison between all models - we regularly find that shielding values from computations employing different, yet reasonable convergence parameters, differ (in terms of their RMSEs) nearly as much as the \shiftmlthree{} predicted shieldings differ from the CSD-test reference calculations.  Instead, we evaluate \shiftmltwo{} and \shiftmlthree{} on 351 structures in our test set that only contain H,C,N,O and are not MD distorted structures (containing 7536 $^1$H, 6546 $^{13}$C,  665 $^{15}$N  and 1311 $^{17}$O test environments) and compare their relative improvements in accuracies against the other models, to give some context to the gains in accuracy we achieve with \shiftmlthree{}. In Table~\ref{tab:comparison_H_C_N_O}, we compare prediction accuracies of the original ShiftML1.0 model, Unzueta's ensemble of GNNs, Liu's MR-3D-DenseNet and compare them with the prediction accuracies of \shiftmltwo{} and \shiftmlthree{} on the H,C,N,O-relaxed sub selection of the CSD-test set.

\begin{table}
    \centering
    \caption{Test set RMSEs (ppm) of the original ShiftML1.0 model (KRR~\cite{paruzzo_chemical_2018}), Unzueta's GNNs (GNN~\cite{unzueta2022low}) and Liu's MR-3D-DenseNet  (DenseNet \cite{liuMultiresolution3DDenseNetChemical2019a}) and Xu's NMRNet (NMRNet~\cite{xuUnifiedBenchmarkFramework2025})  on the CSD-500 test set. We compare these values with the prediction accuracies of \shiftmltwo{} (\shiftmltwoshort{}) and \shiftmlthree{} (\shiftmlthreeshort{}) on a comparable subset of the CSD-test set.  Note that KRR, GNN, DenseNet and NMRNet are evaluated on the CSD-500 test set. As discussed in the text, due to the small overlap of CSD-500 and the CSD-test set, we choose to evaluate \shiftmltwoshort{} and \shiftmlthreeshort{} on a subset of the CSD-test set containing only H,C,N,O structures that are geometry relaxed. (*)~Denotes, that the results of \shiftmltwoshort{} and \shiftmlthreeshort{} were obtained on that sub selection.}
    \begin{tabular}{@{\hskip 2mm}c@{\hskip 2mm}|@{\hskip 4mm}c@{\hskip 4mm}c@{\hskip 4mm}c@{\hskip 4mm}c@{\hskip 2mm}|@{\hskip 2mm}c@{\hskip 4mm}c} \toprule
        Nucleus &KRR& GNN&  DenseNet & NMRNet& \shiftmltwoshort{}*&\shiftmlthreeshort{}*\\ \hline 
         $^{1}$H &0.49&  0.49&  0.37 &0.35&  0.47&0.39\\ 
         $^{13}$C &4.3&  4.06&  3.3 &3.21&  4.07&1.97\\ 
         $^{15}$N &13.3&  9.90&  10.2 &9.45&  12.52&5.71\\ 
         $^{17}$O &17.7&  14.4&  15.3 &13.03&  19.50&9.91\\ \bottomrule
    \end{tabular}

    \label{tab:comparison_H_C_N_O}
\end{table}

\end{document}